\def\mode{1}	
\if 0\mode
\documentclass[journal, twoside, web]{ieeecolor}
\usepackage{lcsys}
\else
\documentclass{IEEEtran}
\IEEEoverridecommandlockouts
\fi

\usepackage[english]{babel}
\usepackage[utf8]{inputenc}
\usepackage[T1]{fontenc}

\usepackage{microtype}
\usepackage{setspace}
\usepackage{blindtext}
\usepackage{siunitx}
\usepackage{comment}
\usepackage{xargs}

\usepackage{titlecaps}
\Addlcwords{or the if a an of for and to -off off in not by via}
\usepackage{graphicx}
\usepackage{color}
\usepackage{epsfig}
\graphicspath{{Images/}}
\usepackage{tikz}
\usepackage{pgfplots,pgfplotstable}
\usetikzlibrary{patterns,fit,matrix}
\usepackage{tikzscale}
\usepackage{scalerel}
\usetikzlibrary{arrows,
	patterns,
	plotmarks,
	svg.path,
	shapes.multipart}
\usepgfplotslibrary{fillbetween}
\pgfplotsset{compat=newest,
	compat/bar nodes=1.8,
	every axis/.append style={
		label style={font=\Large},
		tick label style={font=\large} 
	}
}
\tikzstyle{int}=[draw, fill=black!10, minimum size=5em,thick]
\tikzstyle{init} = [pin edge={to-,thick,black}]
\usepackage{booktabs}

\usepackage{array}

\usepackage{enumitem}
\usepackage{setspace}

\usepackage{amsthm,amsmath,amssymb,amsfonts}
\usepackage{relsize}
\usepackage{nicefrac}
\usepackage{bbm,dsfont}
\usepackage{mathtools}
\usepackage[mathscr]{eucal}
\usepackage[short,c2]{optidef}
\usepackage{dsfont}

\usepackage{soul}

\usepackage[nocompress]{cite}

\usepackage{url}
\usepackage[colorlinks,citecolor=blue,linkcolor=blue]{hyperref}
\usepackage[capitalize,nameinlink]{cleveref}
\newcommand{\orcid}[1]{\href{https://orcid.org/#1}{\includegraphics[scale=0.04]{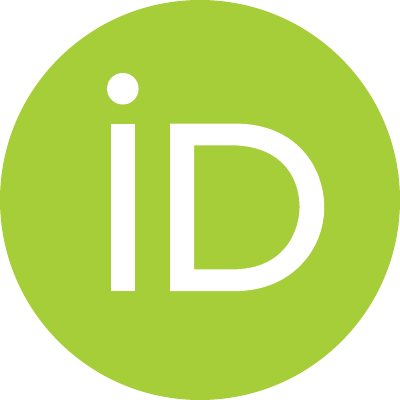}}} 

\newcommand{\eg}{\emph{e.g.,}\xspace}
\newcommand{\ie}{\emph{i.e.,}\xspace}

\newcommand{\Real}[1]{ { {\mathbb R}^{#1} } }

\newcommand{\one}[1][]{\mathds{1}_{#1}}

\newcommand{\e}{\mathrm{e}}

\newcommand{\norm}[2][]{\left\lVert#2\right\rVert_{#1}}

\theoremstyle{remark}
\newtheorem{thm}{Theorem}
\newtheorem{lemma}{Lemma}

\newtheorem{prop}{Proposition}
\newtheorem{ex}{Example}

\newtheorem{rem}{Remark}

\newcommand{\lr}{\left(}
\newcommand{\rr}{\right)}

\newcommand{\agents}{\mathcal{V}}

\newcommand{\state}[2][\@empty]{x%
	\ifx\@empty#1 {_{#2}} \else {_{#2}^{#1}} \fi}

\newcommand{\neigh}[2][\@empty]{\mathcal{N}%
	\ifx\@empty#1 {_{#2}} \else {_{#2}(#1)} \fi}
\newcommand{\neighaug}[2][\@empty]{\neigh[#1]{#2}\cup\{#2\}}

\newcommand{\sumneighaug}[3][\@empty]{\sum_{#3\in\neighaug[#1]{#2}}}
\newcommand{\lam}[2][]{\lambda_{#2}^{#1}}
\newcommand{\lamv}[2][]{\boldsymbol{\lambda}_{#2}^{#1}}

\newcommand{\statess}{x_{\text{ss}}}

\newcommand{\Wleg}{W}
\newcommand{\Waut}[1]{W_{#1}^\text{aut}}
\newcommand{\Win}[1]{W_{#1}^\text{in}}
\newcommand{\prodlam}[3]{\prod_{#1=#2}^{#3}(1-\lam{#1})}

\newcommand{\prodlamfin}[2]{\pi_{#1}^{#2}}

\newcommand{\sigmam}{\sigma_\text{max}}

\addto\captionsenglish{\renewcommand{\figurename}{Fig.}}
\addto\captionsenglish{\renewcommand{\tablename}{TABLE}}

\newcommand{\blue}[1]{{\color{blue}#1}}

\newcommand{\revision}[1]{#1}

\newcommand{\linkToPdf}[1]{\href{#1}{\blue{(pdf)}}}
\newcommand{\linkToPpt}[1]{\href{#1}{\blue{(ppt)}}}
\newcommand{\linkToCode}[1]{\href{#1}{\blue{(code)}}}
\newcommand{\linkToWeb}[1]{\href{#1}{\blue{(web)}}}
\newcommand{\linkToVideo}[1]{\href{#1}{\blue{(video)}}}
\newcommand{\linkToMedia}[1]{\href{#1}{\blue{(media)}}}
\newcommand{\award}[1]{\xspace} 

\if0\mode
\addto\extrasenglish{}

\addto\extrasenglish{}
\else
\addto\extrasenglish{}

\fi

\addto\extrasenglish{}
\addto\extrasenglish{}

\Crefname{prop}{Proposition}{Propositions}

\newcommand{\ptitle}{
	Friedkin-Johnsen Model With\\ Diminishing Competition
}

\if0\mode
\title{\ptitle}
\author{Luca~Ballotta\textsuperscript{\orcid{0000-0002-6521-7142}},
	 \'Aron~V\'ek\'assy\textsuperscript{\orcid{0000-0002-6653-8300}},
	Stephanie~Gil\textsuperscript{\orcid{0000-0002-4951-5350}},
	and~Michal~Yemini\textsuperscript{\orcid{0000-0002-2087-1183}},~\IEEEmembership{Member, IEEE}
    \thanks{This work was supported in part by 
        the EU Horizon Program through the Project TWAIN under Grant 101122194.
    }
	\thanks{Luca Ballotta is with the Delft Center for Systems and Control (DCSC), Delft University of Technology, 2628 CD Delft, Netherlands
		(e-mail: l.ballotta@tudelft.nl).}%
	\thanks{\'Aron V\'ek\'assy and Stephanie Gil  are with the Department of Computer Science, Harvard University, Boston, MA 02138 USA
		(e-mail: avekassy@g.harvard.edu; sgil@seas.harvard.edu).}%
	\thanks{Michal Yemini is with the Faculty of Engineering, Bar-Ilan University, Ramat-Gan 5290002 Israel
		(e-mail: michal.yemini@biu.ac.il).}%
}
\else
\title{\ptitle}
\author{Luca Ballotta,
        \'Aron V\'ek\'assy,
        Stephanie Gil,
        and Michal Yemini
	\thanks{Luca Ballotta is with the Delft Center for Systems and Control (DCSC), Delft University of Technology, Delft, Netherlands
		(e-mail: l.ballotta@tudelft.nl).}%
	\thanks{\'Aron V\'ek\'assy and Stephanie Gil  are with the Department of Computer Science, Harvard University, Boston, MA 02138
		(e-mail: avekassy@g.harvard.edu; sgil@seas.harvard.edu).}%
	\thanks{Michal Yemini is with the Faculty of Engineering, Bar-Ilan University, Ramat-Gan 5290002 Israel
		(e-mail: michal.yemini@biu.ac.il).}%
}
\fi

\begin{document}

    \maketitle

	\if 0\mode
	\pagestyle{empty}
	\thispagestyle{empty}
	\fi
	

\begin{abstract}

	This letter studies the Friedkin-Johnsen (FJ) model with diminishing competition, or stubbornness. 
	The original FJ model assumes \revision{that each agent assigns a constant competition weight to its initial opinion.}
	\revision{In contrast,}
    we investigate the effect of diminishing competition on the convergence point and speed of the FJ dynamics.
	We prove that,
	if the competition is uniform across agents and vanishes asymptotically,
	the convergence point coincides with the nominal consensus reached with no competition. 
	However, the diminishing competition slows down convergence according to its own rate of decay.
	We study this phenomenon analytically and provide upper and lower bounds on the convergence rate.
	Further,
    if competition is not uniform across \revision{agents},
	we show that the convergence point may not coincide with the nominal consensus point. 
	Finally, we evaluate our analytical insights numerically.

	\if 0\mode
	\begin{IEEEkeywords}
		Friedkin-Johnsen model, 
		diminishing competition, 
		consensus,
		convergence, 
		convergence rate.
	\end{IEEEkeywords}
	\fi
	
\end{abstract}

\section{Introduction}

\if0\mode
\IEEEPARstart{T}{he recent} 
\else
The recent
\fi
decades have witnessed a substantial intertwining between systems theory and opinion dynamics over social networks.
Parallel to the development of mathematical models for the exchange and aggregation of opinions among individuals,
tools from systems theory have proven useful to analytically understand their behavior.
Among the models aiming to explain social interactions,
the ``opinion pooling'' proposed by DeGroot~\cite{degroot} and the Friedkin-Johnsen (FJ) model~\cite{FJdynamics} have gained a great deal of attention in the control community.
The former relates to a significant body of literature on distributed consensus,
widely used in distributed and multi-agent control problems and pioneered by works such as~\cite{tsitsiklis1984problems,Jadbabaie03tac-coordinationMobileAgents,XIAO200465}.
Relevant works on the FJ model analyze convergence with constant and time-varying parameters~\cite{Proskurnikov17ifacwc-timeVaryingFJ,Parsegov17tac-multidimensionalFJ,Zhou22infosci-FJincreasingStubbornness},
multidimensional extensions~\cite{Parsegov17tac-multidimensionalFJ,Zhou22infosci-FJincreasingStubbornness},
interactions over random graphs~\cite{Wang24arxiv-randomGraphsFJ},
and mixed dynamics with one or two timescales~\cite{Disaro24automatica-homophilyFJ,Jia17jco-deGrootFriedkin,Bernardo21sciadv-parisAgreementFJ}.
However,
because the FJ model,
similarly to other models of opinion dynamics,
captures prejudice and disagreement that are natural features of human behavior,
it has been underutilized in control problems, 
which often consider coordination among cooperative agents.

Nonetheless,
a few recent works propose tools of opinion modeling for distributed control systems.
In fact,
the primary motivation for this letter spawns from previous works~\cite{ballotta24tac-compColl,ballottaCDC22compColl,acc} that use the FJ model to enhance resilience of consensus tasks to adversaries in multi-agent systems~\cite{6481629}.
In particular,
the stubbornness parameter of the FJ model is used in~\cite{ballotta24tac-compColl,ballottaCDC22compColl} to introduce controlled competition in the system,
mitigating the influence of unknown malicious agents.
Additionally,
along the lines of trust-based resilient consensus enabled by the physical communication channel in cyberphysical systems~\cite{Yemini22tro-resilienceConsensusTrust},
the recent paper~\cite{acc} relaxes the model in~\cite{ballotta24tac-compColl} by letting the competition parameter vanish,
which allows the agents to reach a consensus whose deviation from nominal depends on how fast they identify adversaries in the system.

However,
while~\cite{ballotta24tac-compColl,acc} focus on interactions between ``legitimate'' and malicious agents,
the nominal case with no adversaries is overlooked,
and in particular~\cite{acc} does not discuss the consensus reached in that scenario.
In fact,
despite the rich literature on the FJ model,
the case where the competition parameter vanishes has not been studied.
This may indeed be unrealistic in social systems where individuals do not reasonably forget their initial opinions.
However,
it may be useful in an engineering setting where the behavior is imposed by a system designer,
such as to be robust to adversaries as proposed in~\cite{ballotta24tac-compColl,acc}
\revision{or to mitigate undesired large oscillations of the agents' states caused by the consensus protocol,
	thus enhancing the safety or lifespan of physical components via a simple algorithm with no constraints.}
We fill this gap and systematically study the FJ model with vanishing competition.

\subsubsection*{Contribution}
The main contribution of this letter is threefold.
First,
we show that the FJ model with vanishing competition reaches the same steady state of the ``nominal'' consensus protocol (DeGroot model) that uses the same weights.
This not only broadens the scope of the existing literature on the FJ model,
but mostly reinforces the motivation of using it for resilient coordination as done in~\cite{acc} since,
with this result,
the agents are guaranteed to reach the nominal consensus if no adversary is present.
Second,
we quantify how fast the trajectory approaches the steady state.
Our results reveal that the convergence rate is dominated by the slowest between the nominal consensus dynamics and the vanishing competition parameter.
Finally,
we prove a relevant negative result,
\ie convergence to the nominal consensus point may not occur if the competition parameter,
albeit vanishing,
is non-uniform across the agents.
This demands attention when deploying the protocol for multi-agent systems and in particular may limit the performance of a fully decentralized design.

\section{Setup}
    		\revision{

\subsection{Notation and Mathematical Preliminaries}\label{sec:notation}
A matrix $W$ is nonnegative (positive),
denoted by $W\ge0$ ($W>0$),
if all its elements $W_{ij}$ are nonnegative (positive).
Additionally, a matrix $W$ is (row-)stochastic if $W\ge0$ and $W\one = \one$,
where $\one$ is the vector of all ones.
Matrix $W$ is primitive if there exists $k\in\mathbb{N}$ s.t. $W^k > 0$.
The Perron eigenvector $v$ of stochastic primitive matrix $W$ is the unique stochastic vector satisfying $W^\top v = v$.
The notation $a\not\parallel b$ means that vectors $a\in\Real{n}$ and $b\in\Real{n}$ are not parallel,
or equivalently that their cross product $a \times b \neq 0$.}

\subsection{System Model}\label{sec:system-model}

We consider $N$ agents equipped with scalar-valued states.
\revision{Vector-valued states do not change the results but make notation heavy.}
We denote the state of agent $i$ at time $t$ by $\state[i]{t}\in\Real{}$,
with $i \in \agents \doteq \{1,\dots,N\}$,
and the vector with all stacked states by $\state{t}{}\in\Real{N}$.
The agents exchange their state values through a fixed communication network,
modeled as a graph $\mathcal{G} = (\agents, \mathcal{E})$.
Each element $(i,j) \in \mathcal{E}$ indicates that agent $j$ can transmit data to agent $i$ through a direct communication link.

Let $\neigh{i}\in\agents$ denote the set of neighbors of agent $i$ in the communication network $\mathcal{G}$,
\ie $\neigh{i} \doteq \{j\in\agents : (i,j)\in\mathcal{E}\}$.
The agents weight information coming from neighbors through the nonnegative matrix $\Wleg$,
such that $\Wleg_{ij} > 0$ if and only if $j\in\neigh{i}\cup\{i\}$.
We assume that $\Wleg$ is stochastic and primitive.

\subsection{FJ Model With Uniform Time-Varying Competition}\label{sec:resilient-protocol}

In this work,
we study the following protocol implemented by each agent $i\in\agents$ for $t\ge0$:
\begin{equation}\label{eq:update-rule-regular}
	\state[i]{t+1} =  \lr1-\lam{t}\rr\sumneighaug{i}{j}W_{ij}\state[j]{t} + \lam{t}\state[i]{0}.
\end{equation}
The scalar time-varying parameter $\lam{t}\in[0,1]$ represents \textit{competition} or \textit{stubbornness} of the agents.
There are two extreme cases for this model.
When $\lambda_t\equiv0$, 
then~\eqref{eq:update-rule-regular} reduces to the consensus protocol or DeGroot model~\cite{degroot,XIAO200465} that leads to the consensus $\lim_{t\rightarrow\infty}\state{t}=\one v^\top\state{0}$.
Also, when $\lambda_t\equiv\lambda>0$,
then~\eqref{eq:update-rule-regular} is the standard FJ model~\cite{FJdynamics},
which does not lead to consensus except for trivial cases~\cite{Proskurnikov17ifacwc-timeVaryingFJ}.

Our main motivation to study the protocol~\eqref{eq:update-rule-regular} stems from previous works~\cite{ballotta24tac-compColl,acc} on resilient consensus,
where the nominal consensus protocol $\state{t+1}=W\state{t}$ may be disrupted by unknown agents~\cite{6481629,Pasqualetti12tac-resilientConsensus,Yemini22tro-resilienceConsensusTrust}.
The intuition behind using~\eqref{eq:update-rule-regular} with constant $\lam{t}\equiv\lambda$ in~\cite{ballotta24tac-compColl} is to prevent normal agents be overly influenced by their neighbors without knowing if these are malicious,
thus sacrificing consensus for a safe trajectory.
To recover a steady-state consensus,
the paper~\cite{acc} uses~\eqref{eq:update-rule-regular} with a vanishing competition parameter $\lam{t}$ and time-varying weights $W_{ij}(t)$,
whereby the normal agents detect adversaries through exogenous physics-based information -- and tune weights accordingly -- when $\lam{t}$ becomes small.
However,
the convergence in the nominal adversary-free case was not investigated.
Hence, in this work our main focus is twofold: 
1) understanding the limit point of the dynamics~\eqref{eq:update-rule-regular} and its relation to the two extreme cases of $\lam{t}\equiv0$ and $\lam{t}\equiv1$;
2) quantifying the effect of the sequence $\{\lambda_t\}_{t\ge0}$ on its convergence rate.

Before presenting our results, we introduce a few helpful \revision{facts}.
First, in compact vector form,
the update rule~\eqref{eq:update-rule-regular} reads
\begin{equation}\label{eq:fj}\tag{FJ}
	\state{t+1} = (1-\lam{t})\Wleg\state{t} + \lam{t}\state{0} .
\end{equation}
Additionally,
for $s,t\in\Real{}\cup\{\infty\}$,
we define the scalar
\revision{\begin{equation}\label{eq:prod-lam}
	\prodlamfin{t}{s} \doteq \begin{cases}
		 \prodlam{k}{t}{s},	& s \ge t\\
		 1,								& s < t
	\end{cases}
\end{equation}
with $\prodlamfin{t}{\infty} \doteq \lim_{s\to\infty}\prodlamfin{t}{s}$.}
\revision{Consequently},
for an initial condition $\state{0}$,
the state $\state{t}$ can be written as the linear transformation
\begin{equation}\label{eq:state-components}
	\state{t} = \lr\Waut{t} + \Win{t}\rr\state{0}
\end{equation}
where the two transition matrices associated with the autonomous consensus dynamics and \revision{the input signal $\lam{t}\state{0}$ respectively are derived as}
\begin{align}
	\Waut{t}	&= \prodlamfin{0}{t-1}W^t \label{eq:W-aut}\\
	\Win{t}		&= \sum_{k=0}^{t-1} \prodlamfin{k+1}{t-1}W^{t-1-k}\lam{k}. \label{eq:W-in}
\end{align}
The rest of this letter delves into the convergence of~\eqref{eq:fj},
focusing on the comparison with the consensus protocol.

\section{Convergence Analysis}\label{sec:performance-analysis}

This section presents our main results on convergence properties of the protocol~\eqref{eq:fj}.
First,
\autoref{sec:convergence} establishes that~\eqref{eq:fj} converges to the same equilibrium reached by the consensus protocol with matrix $W$.
Then,
\autoref{sec:convergence-rate} bounds its convergence rate,
\ie how fast the steady state is approached,
highlighting the role of the competition parameter $\lam{t}$.

\subsection{Convergence to Consensus}\label{sec:convergence}

Here,
we step up the intuition of effectively using competition for resilience by showing that protocol~\eqref{eq:fj} makes the agents ultimately agree on the nominal consensus value.

While it is intuitive that the autonomous dynamics-related term $\Waut{t}\state{0}$ in~\eqref{eq:state-components} reaches a consensus,
the next nontrivial results shows that also the input-related term in~\eqref{eq:state-components} achieves a steady-state consensus if $\lam{t}$ asymptotically vanishes.

\begin{lemma}\label{lemma:convergence-input}
	If $\lim_{t\rightarrow\infty}\lambda_t=0$,
	then there exists $y\in\Real{N}$ such that 
	\begin{equation}\label{eq:convergence-input}
		\lim_{t\to\infty}\Win{t}=\one y^\top .
	\end{equation}
	\begin{proof}
		See \cref{app:input-consensus}.
	\end{proof}
\end{lemma}

\Cref{lemma:convergence-input} leads us to our first main result,
which establishes that the steady state reached through~\eqref{eq:fj} is the nominal consensus obtained through the consensus protocol.
Notably,
this cannot be concluded from previous work~\cite{Proskurnikov17ifacwc-timeVaryingFJ} on time-varying FJ model,
where the competition parameter is not allowed to vanish and consensus happens only in the trivial case where the initial condition $\state{0}$ is already a consensus.

\begin{thm}[Protocol~\eqref{eq:fj} achieves the nominal consensus]\label{thm:consensus}
	Let $v$ be the Perron eigenvector of $\Wleg$ and $\statess \doteq v^\top\state{0}\in\Real{}$.
	If $\state{0} \not\parallel \one$,
	then the following two statements are equivalent:
	\begin{enumerate}[labelindent*=0pt]
		\item $\lim_{t\rightarrow\infty}\lambda_t=0$;
		\item $\lim_{t\rightarrow\infty}\state{t}= \statess\one$.
	\end{enumerate}
	\begin{proof}
		See \cref{app:consensus}.
	\end{proof}
\end{thm}

In words,
\Cref{thm:consensus} states that the competition parameter $\lam{t}$ in~\eqref{eq:fj} does not affect the steady state as long as it asymptotically vanishes.
This not only recovers the intuition that the consensus dynamics in~\eqref{eq:fj} dominates when $\lam{t}$ is small,
but also proves that the state $\state{t}$ approaches the same consensus regardless of the specific sequence $\{\lam{t}\}_{t\ge0}$.

Nonetheless,
the convergence speed of~\eqref{eq:fj} heavily depends on how fast the competition decays,
which is addressed next.

\subsection{Convergence Rate}\label{sec:convergence-rate}

We now assess how fast the protocol~\eqref{eq:fj} converges.
In view of \Cref{thm:consensus},
we assume $\lim_{t\to\infty}\lam{t} = 0$.
Intuitively,
if the competition parameter $\lam{t}$ decays slowly,
it slows down convergence because the agents stick close to their initial conditions for long time.
Interpreting~\eqref{eq:fj} as a (time-varying) exponential moving-average filter,
increasing $\lam{t}$ reduces the cutoff frequency and attenuates the high frequencies~\cite{nau2016statistical}.

To estimate the convergence speed of~\eqref{eq:fj},
we proceed to bound the function $\rho(t)$ defined as
\begin{equation}\label{eq:convergence-rate}
	\rho(t) \doteq \sup_{\state{0}\not\parallel\one}\dfrac{\norm[2]{\state{t} - \statess\one}}{\norm[2]{\state{0} - \statess\one}}.
\end{equation}
Requiring $\state{0}\not\parallel\one$ in~\eqref{eq:convergence-rate},
\ie that $\state{0}$ is not a consensus,
is not restrictive because \revision{if $\state{0} = \alpha\one$ for some  $\alpha\in\Real{}$,
then the protocol~\eqref{eq:fj} generates the sequence $\state{t} \equiv \alpha\one$,
thus it yields trivial instantaneous convergence which we preclude.}

Because the input $\lam{t}\state{0}$ is continually injected,
standard convergence tools for consensus cannot be used,
preventing us to establish geometric convergence based on norms of the weight matrix.
Moreover,
standard results on the convergence speed of the FJ model~\cite{Proskurnikov17ifacwc-timeVaryingFJ} and on time-varying consensus~\cite{Blondel05cdc-consensusConvergence,Xiao06automatica-consensusMetropolisWeights} are inadequate because they assume either zero or strictly positive parameters,
whereas $\lam{t}$ goes to zero.

Toward the presentation of the convergence speed results only,
we assume that $\Wleg$ is doubly stochastic so we can derive nontrivial bounds based on the singular value decomposition.
Consequently, let $\sigmam\in(0,1)$ denote the second largest singular value of $\Wleg$.
Recall that $\prodlamfin{s}{t}$ is defined in~\eqref{eq:prod-lam},
and define
\begin{multline}\label{eq:convergence-rate-upper-bound}
	\bar{\rho}(t)\doteq
	\prodlamfin{0}{t-1}\lr \sigmam^{t} +  1-\prodlamfin{t}{\infty} \rr \\
	+\sum_{k=0}^{t-1}\prodlamfin{k+1}{t-1}\lam{k}\lr\sigmam^{t-1-k} +  1-\prodlamfin{t}{\infty}\rr 
	+ \sum_{k=t}^{\infty}\prodlamfin{k+1}{\infty}\lam{k}.
\end{multline}
We first introduce an upper bound on the convergence rate.

\begin{prop}\label{prop:convergence-rate-upper-bound}
	For every doubly stochastic matrix $W$ and \revision{$\state{0}\not\parallel\one$},
	it holds
	\begin{equation}
		\label{eq:convergence-rate-bound}
		\dfrac{\norm[2]{\state{t} - \statess\one}}{\norm[2]{\state{0} - \statess\one}} \le \bar{\rho}(t) \qquad \forall \; t\ge1.
	\end{equation}
	\begin{proof}
		See \cref{ass:proof-convergence-rate-upper-bound}.
	\end{proof}
\end{prop}

The bound $\bar{\rho}(t)$ depends not only on $\sigmam^{t}$ but also on all powers $\sigmam^{t-1-k}$ of the dominant singular value,
making convergence slower.
Also,
it holds $\lim_{t\to\infty}\bar{\rho}(t) = 0$ since
\begin{equation}\label{eq:limits-rho}
    \begin{aligned}
        \lim_{t\to\infty} \prodlamfin{t}{\infty} &= 1\\
        \lim_{t\to\infty} \lam{k}\sigmam^{t-1-k} &= 0 \qquad \forall k = 0,\dots,t-1.
    \end{aligned}
\end{equation}
We next introduce a lower bound in the spirit of \cite{1638541}.
Let us define
\begin{equation}\label{eq:conv-rate-lower-bound}
	\underline{\rho}(t)\doteq\prodlamfin{0}{t-1} \sigmam^t +  \sum_{k=0}^{t-1}\prodlamfin{k+1}{t-1}\lam{k}\sigmam^{t-1-k}.
\end{equation}

\begin{prop}\label{prop:convergence-rate-lower-bound}
	For every doubly stochastic matrix $W$,
	there exists $\state{0}$ such that
	\begin{equation}
		\dfrac{\norm[2]{\state{t} - \statess\one}}{\norm[2]{\state{0} - \statess\one}} \ge \underline{\rho}(t) \qquad \forall\;t\ge1.
	\end{equation}
	\begin{proof}
		See \cref{ass:proof-convergence-rate-lower-bound}.
	\end{proof}
\end{prop}

\revision{The lower bound $\underline{\rho}(t)$ has the same dependency on $\sigmam$, and hence on the communication matrix $W$,
of the upper bound $\bar{\rho}(t)$,
and goes to zero by the same argument in~\eqref{eq:limits-rho}.}
Putting together the two bounds readily yields our second main result.

\begin{thm}[Convergence rate of~\eqref{eq:fj}]\label{thm:convergence-rate}
	For every stochastic matrix $W$,
	the convergence rate defined in~\eqref{eq:convergence-rate} is bounded as
	\begin{equation}
		\underline{\rho}(t) \leq \rho(t) \leq \bar{\rho}(t) \qquad \forall\;t\ge1.
	\end{equation}
	\begin{proof}
		It follows immediately by applying \cref{prop:convergence-rate-upper-bound,prop:convergence-rate-lower-bound} to the definition of convergence rate in~\eqref{eq:convergence-rate}.
	\end{proof}
\end{thm}

\Cref{thm:convergence-rate} quite precisely quantifies how fast the protocol converges.
The input $\lam{t}\state{0}$ prevents geometric convergence in general.
\revision{In fact,
both bounds $\bar{\rho}(t)$ in~\eqref{eq:convergence-rate-upper-bound} and $\underline{\rho}(t)$ in~\eqref{eq:conv-rate-lower-bound} include all powers of $\sigmam$ up to the current time and specifically the products $\prodlamfin{k+1}{t-1}\lam{k}\sigmam^{t-1-k}$,
which embed the decay rate of $\lam{t}$ and quantify how slow~\eqref{eq:fj} is compared to the exponential convergence rate $\sigmam^t$ of the consensus dynamics.}
The gap $\bar{\rho}(t) - \underline{\rho}(t)$ between upper and lower bounds does not depend on $\sigmam$ -- and thus on $W$ -- but only on $\{\lam{t}\}_{t\ge0}$ and amounts to
\begin{equation}\label{eq:bounds-diff}
    \prodlamfin{0}{t-1} - \prodlamfin{0}{\infty} + \sum_{k=0}^{t-1}\lr\prodlamfin{k+1}{t-1} - \prodlamfin{k+1}{\infty}\rr\lam{k} + \sum_{k=t}^\infty\prodlamfin{k+1}{\infty}\lam{k}.
\end{equation}
The quantity in~\eqref{eq:bounds-diff} vanishes with $t$,
showing that asymptotically the two bounds overlap,
and also,
for every $t$, it is small if $\lam{t}$ decays fast.
However,
while the rate $\rho(t)$ is precisely quantified in these two regimes,
they represent different behaviors.
On the one hand,
the convergence speed is generally ruled by both $W$ and $\lam{t}$ for large $t$;
on the other hand,
if $\lam{t}$ decays very fast,
both bounds $\bar{\rho}(t)$ and $\underline{\rho}(t)$ approach the exponential rate $\sigmam^t$ \revision{for every $t$},
which is indeed intuitive because~\eqref{eq:fj} essentially reduces to the consensus dynamics.

\begin{rem}[Slow convergence for resilience]
    Paper~\cite{acc} precisely leverages the slower convergence of~\eqref{eq:fj} for resilient consensus.
    Indeed,
    the normal agents can identify the adversaries before deviating too much from the initial condition,
    mitigating malicious messages erroneously considered trustworthy.
\end{rem}

\begin{ex}
    Consider the sequence $\lambda_t = \frac{1}{t+1}$.
    \revision{Because the series of its partial sums diverges,
    the products $\prodlamfin{s}{t}$ ``diverge to zero'',
    \ie $\prodlamfin{s}{\infty}=0$ for finite $s$~\cite{infProductConvergence}}.
    Hence,
    the autonomous term $\Waut{t}\state{0}$ vanishes and the steady state is reached through only the input sequence.
    Also,
    we can explicitly compute $\prodlamfin{s}{t}=\frac{s}{t+1}$ and the two bounds on the convergence rate read
    \begin{equation}
        \bar{\rho}(t) = \sum_{k=0}^{t-1} \frac{\sigmam^{t-1-k}+1}{t}
        \qquad
        \underline{\rho}(t) = \sum_{k=0}^{t-1} \frac{\sigmam^{t-1-k}}{t}.
    \end{equation}
    This reveals that $\rho(t) \in O(\nicefrac{1}{t})$,
    namely the convergence rate is dominated by the decay rate of the competition parameter.
    This is a significant reduction in the convergence speed of the nominal consensus protocol,
    where $\lambda_t\equiv0$,
    that guarantees a convergence rate of exponential order $\sigmam^{t}$.
\end{ex}

\subsection{Non-Uniform Competition: A Negative Result}
\label{sec:nonuniform-comp}

While so far we have considered $\lam{t}$ identical for all agents in~\eqref{eq:update-rule-regular},
the competition parameter of the standard FJ model may differ across the agents.
Dropping the uniform assumption and considering the vector-valued parameter $\lamv{t}\in\Real{N}$ with $i$th element $\lam[i]{t}\in[0,1]$ changes \eqref{eq:fj} to the more general form
\begin{equation}\label{eq:fj_nonuniform}
	\state{t+1} = \mathrm{diag}(\one-\lamv{t})\Wleg\state{t} + \mathrm{diag}(\lamv{t})\state{0}.
\end{equation}
The following proposition shows that,
under this modified protocol, 
the analogous result of \cref{thm:consensus} does not hold.

\begin{prop}\label{cor:nonuniform}
	If the agents run \eqref{eq:fj_nonuniform}, then for any stochastic matrix $W$ and initial state $\state{0} \not\parallel \one$ there exists a sequence $\{\lamv{t}\}_{t\ge0}$ with $\lim_{t\rightarrow\infty}\lamv{t}=0$ such that $\lim_{t\rightarrow\infty}\state{t}\neq\statess\one$.
    \begin{proof}
        See \Cref{app:nonuniform}.
    \end{proof}
 \end{prop}

\cref{cor:nonuniform} highlights important limitations of the protocol~\eqref{eq:fj_nonuniform} as compared to~\eqref{eq:fj}.
For example,
while the consensus protocol can be implemented in a distributed way,
\eg letting the agents independently set uniform or Metropolis weights~\cite{Xiao06automatica-consensusMetropolisWeights},
the same does not hold for protocol~\eqref{eq:fj_nonuniform} because convergence to the nominal consensus is not guaranteed if the agents use different competition parameters.
Moreover,
centrally setting the same parameter $\lambda$ for all agents is reasonable in some cases,
such as a team of robots programmed by a company,
but impractical in other cases,
such  as a large-scale power plant managed by several independent stakeholders.
Therefore,
a compelling direction of future research is to let agents independently choose their competition parameters $\lambda_i$ with guarantees on the final consensus point.

\section{Numerical Results}
\label{sec:sims}

\begin{figure} 
    \centering
    \includegraphics[width=1\linewidth]{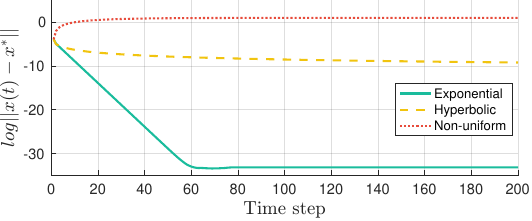}
    \caption{Average logarithmic distance of $20$ agents from the nominal consensus value using the three different competition parameters $\lambda_t$ described in \autoref{sec:sims}.}
    \label{fig:simulations}
\end{figure}

To complement the theoretical results, we provide a simulation that illustrates the effect the different choices of competition parameters have on the convergence of protocol~\eqref{eq:fj}. 
This is shown in \autoref{fig:simulations}. 
We simulated $20$ agents interconnected along an Erdős-Rényi random graph with $p=0.1$. 
Their initial values $\state[i]{0}$ were chosen uniformly random from $[0, 5]$. 
The cases labeled as ``exponential'' and ``hyperbolic'' were run according to \eqref{eq:fj} with $\lam{t}=\e^{-0.5t}$ and $\lam{t}=(t+1)^{-1}$, respectively.
Meanwhile, in the case labeled as ``non-uniform'' the agents followed \eqref{eq:fj_nonuniform} with the competition parameters described in \Cref{app:nonuniform} with the choice $t^* = 100$. 
According to the theory,
in the exponential and hyperbolic cases the agents converge to the nominal consensus value, albeit with different convergence rates, while in the non-uniform case the process deviates.

\section{Conclusion}

We have studied an FJ model with time-varying competition parameter $\lam{t}$,
establishing convergence to the steady state $\one v^\top\state{0}$ of the corresponding consensus protocol if and only if $\lam{t}$ vanishes.
This equivalence fails if agents use heterogeneous parameters $\lam[i]{t}$.
We have also derived bounds on the convergence rate that quantify how the sequence $\{\lam{t}\}_{t\ge0}$ slows down convergence.
Numerical tests exemplify the theory results.

    \appendix[Proofs of Results]
    \numberwithin{equation}{subsection}
    

\subsection{Proof of \Cref{lemma:convergence-input}}
\label{app:input-consensus}

We prove that
\begin{equation}\label{eq:limit-input}
	\lim_{t\to\infty} \Win{t} = \one v^\top\sum_{k=0}^\infty \prodlamfin{k+1}{\infty}\lam{k}.
\end{equation}
The intuition behind this limit is that,
for $k\to\infty$,
the distance from consensus dynamics vanishes through $\lam{k}\to0$.
Importantly,
the limit of the scalar coefficient in the RHS of~\eqref{eq:limit-input} exists finite and is positive for all decreasing sequences $\{\lam{t}\}_{t\ge0}$,
which we show as a by-product of the proof in \Cref{app:consensus}.
The statement above can be proven as
\begin{equation}\label{eq:limit}
	\lim_{t\to\infty}\norm{\Win{t} - \one v^\top\sum_{k=0}^\infty \prodlamfin{k+1}{\infty}\lam{k}} = 0.
\end{equation}
Splitting the infinite summation and product in the limit so as to take apart the summation from $0$ to $t$ and the product from $k+1$ to $t$,
respectively,
we rewrite~\eqref{eq:limit} as
\begin{multline}\label{eq:limit-1}
	\lim_{t\to\infty}\norm{\sum_{k=0}^t \prodlamfin{k+1}{t}\lr\Wleg^{t-k} - \one v^\top\prodlamfin{t+1}{\infty}\rr \lam{k} \right. \\
		\left. - \one v^\top\sum_{k=t+1}^{\infty}\prodlamfin{k+1}{\infty}\lam{k}} = 0.
\end{multline}
The triangle inequality and the Sandwich theorem yield the following upper bound to the limit in the LHS of~\eqref{eq:limit-1}:
\begin{multline}\label{eq:limit-tr-ineq}
	\lim_{t\to\infty}\sum_{k=0}^t\prodlamfin{k+1}{t}\norm{\Wleg^{t-k} - \one v^\top\prodlamfin{t+1}{\infty}}\lam{k} \\
	+ \norm{\one v^\top}\lim_{t\to\infty}\sum_{k=t+1}^{\infty}\prodlamfin{k+1}{\infty}\lam{k}.
\end{multline}
The second limit in~\eqref{eq:limit-tr-ineq} is zero because all addends vanish.
The sum of nonnegative terms in the first limit converges to zero if and only if each term converges to zero.
For all values of $t$ and $k$,
it holds $\prodlamfin{k+1}{t} \in (0,1)$.
Let
\begin{equation}
	\alpha(t,k) \doteq \norm{\Wleg^{t-k} -\one v^\top\prodlamfin{t+1}{\infty}}\lam{k}.
\end{equation}
We next compute the limit of $\alpha(t,k)$ as $t\to\infty$.
According to the difference $t-k$,
we consider two asymptotic cases.
\begin{description}
	\item[\boldmath $\lim_{t\to\infty} (t-k) = \infty$:] It holds $\prodlamfin{t+1}{\infty}\to1$ and
	\begin{equation}
		\lim_{t\to\infty} \alpha(t,k) = \norm{\Wleg^{\infty} - \one v^\top}\lam{k} = 0 \cdot \lam{k}  = 0.
	\end{equation}
	\item[\boldmath $\lim_{t\to\infty,k\to\infty} (t-k) = \alpha \in \mathbb{N}$:] It holds $\prodlamfin{t+1}{\infty}\to1$ and
	\begin{equation}
		\lim_{\substack{t\to\infty\\k\to\infty}} \alpha(t,k) = \norm{\Wleg^\alpha - \one v^\top} \lim_{k\to\infty}\lam{k} = \kappa \cdot 0 = 0
	\end{equation}
	where $\norm{\Wleg^\alpha - \one v^\top} = \kappa \in \Real{}$.
\end{description}
Because $\alpha(t,k)\to0$ for all $k\le t$,
the limit in~\eqref{eq:limit-tr-ineq} is zero,
which implies~\eqref{eq:limit-1} and in turn the claim~\eqref{eq:convergence-input}.

\subsection{Proof of \cref{thm:consensus}}
\label{app:consensus}

\paragraph*{1) implies 2)}
From~\eqref{eq:state-components} with~\eqref{eq:W-aut}--\eqref{eq:W-in}, it follows
\begin{equation}\label{eq:lim-state}
	\begin{aligned}
		\lim_{t\rightarrow\infty}\state{t} 
		&=\lim_{t\to\infty} \lr\Waut{t} + \Win{t}\rr\state{0}
		=\lr\one v^\top \prodlamfin{0}{\infty} + \lim_{t\to\infty}\Win{t}\rr\state{0}\\
		&=\one v^\top \lr\prodlamfin{0}{\infty} + \sum_{k=0}^{\infty}\prodlamfin{k+1}{\infty}\lam{k}\rr \state{0}
	\end{aligned}
\end{equation}
where the last equality uses~\eqref{eq:limit-input}.
We now prove that the factor between brackets in the RHS of the last equality in~\eqref{eq:lim-state} is $1$.
This can be done by induction.
We prove that
\begin{equation}\label{eq:induction-1}
	\prodlamfin{0}{t}  + \sum_{k=0}^t \prodlamfin{k+1}{t}\lam{k} = 1 \qquad \forall t\ge0.
\end{equation}
\begin{description}[leftmargin=0cm]
	\item[Base step] For $t = 0$,
	it holds 
	\begin{equation}
		\prodlamfin{0}{0}  + \sum_{k=0}^0 \prodlamfin{k+1}{0}\lam{k} = 1 - \lam{0} + 1 \cdot \lam{0} = 1.
	\end{equation}
	\item[Induction step] Let the claim hold for $t-1$.
	For $t$,
	it holds
	\begin{align}
		\begin{split}
			\prodlamfin{0}{t} + \sum_{k=0}^{t} \prodlamfin{k+1}{t}\lam{k} 
			&=(1-\lam{t})\prodlamfin{0}{t-1} + \sum_{k=0}^{t-1} \prodlamfin{k+1}{t}\lam{k} + 1 \cdot \lam{t}\\
			&=(1-\lam{t})\lr \prodlamfin{0}{t-1} + \sum_{k=0}^{t-1} \prodlamfin{k+1}{t-1}\lam{k}\rr + \lam{t}\\
			&=(1-\lam{t}) \cdot 1 + \lam{t} = 1.
		\end{split}
	\end{align}
\end{description}
From~\eqref{eq:lim-state} and~\eqref{eq:induction-1},
it immediately follows that
\begin{equation}
	\lim_{t\rightarrow\infty}\state{t} = \one v^\top \state{0} = \statess\one.
\end{equation}

\paragraph*{2) implies 1)}
\revision{We prove this direction by contradiction. 
	First,
	let $\lambda\doteq\lim_{t\to\infty}\lam{t}>0$.
	If $\lim_{t\to\infty}\state{t} = \statess\one$,
	then it holds $\statess\one = \lambda\state{0} + (1-\lambda)W\statess\one = \lambda\state{0} + (1-\lambda)\statess\one$.
	If $\lambda > 0$,
	this happens if and only if $\state{0} = \statess\one$,
	against the hypothesis $\state{0}\not\parallel\one$.
    Second,
	let $\nexists \lim_{t\to\infty}\lam{t}$,
	then $\state{t}\to\statess\one$ if and only if $\lam{t}\state{0} + (1-\lam{t})\state{t} \to\statess\one$.
	The latter is true if,
	$\forall \epsilon>0$,
	$\exists T$ s.t. $\norm{\lam{t}\state{0} + (1-\lam{t})\state{t} - \statess\one}<\epsilon \; \forall t\ge T$. 
    The triangle inequality yields $\norm{\lam{t}\state{0} + (1-\lam{t})\state{t} - \statess\one} \geq \lam{t}\norm{\state{0} - \state{t}} - \norm{\state{t} - \statess\one}$.
    From \textit{2)},
    $\forall\epsilon'>0$ $\exists T'$ s.t. $\norm{\state{t} - \statess\one}<\epsilon'$ for all $t\ge T'$.
    Thus, for $t\ge\max\{T,T'\}$,
    it holds $\lam{t}\norm{\state{0} - \state{t}} - \norm{\state{t} - \statess\one}> \lam{t}\norm{\state{0} - \state{t}} - \epsilon'$.
    Hence,
    it must be $\lam{t}\norm{\state{0} - \state{t}} < \epsilon' \; \forall t\ge\max\{T,T'\}$,
    which holds true if and only if $\state{t}\to\state{0}$,
    against the hypothesis $\state{0}\not\parallel\one$.
    }

\subsection{Proof of \Cref{prop:convergence-rate-upper-bound}}
\label{ass:proof-convergence-rate-upper-bound}

By definition,
the mismatch between the state contribution at time $t$ and the final (asymptotic) value is
\begin{equation}
	\begin{aligned}
		\state{t} -\statess \one 	&= (\Waut{t} + \Win{t})\state{0} - (\Waut{\infty} + \Win{\infty})\state{0}
	\end{aligned}
\end{equation}
where,
since $W$ is stochastic and primitive and $\lim_{t\to\infty}\lam{t}=0$,
from~\eqref{eq:W-aut} and~\eqref{eq:limit-input} the limits of $\Waut{t}$ and $\Win{t}$ are
\begin{equation}\label{eq:limits-W}
	\Waut{\infty} = \one v^\top\prodlamfin{0}{\infty}, \qquad 
	\Win{\infty} = \one v^\top\sum_{k=0}^\infty \prodlamfin{k+1}{\infty}\lam{k}.
\end{equation}
Because $(\Waut{t} + \Win{t})\one = \one \ \forall t\ge0$,
it holds
\begin{equation}\label{eq:state-mismatch-leg}
	\state{t} -\statess \one = (\Waut{t} + \Win{t} - \Waut{\infty} - \Win{\infty}) (\state{0} - \statess\one).
\end{equation}
The Schwartz and triangle inequalities respectively yield
\begin{gather}
	\norm[2]{\state{t} - \statess\one}	\! \le \!\norm[2]{\Waut{t} + \Win{t} - \Waut{\infty} - \Win{\infty}}\!\norm[2]{\state{0} - \statess\one} \label{eq:state-norm-ineq}\\
	\begin{multlined}\label{eq:state-mismatch-leg-norm}
		\norm[2]{\Waut{t} + \Win{t} - \Waut{\infty} - \Win{\infty}} \le \norm[2]{\Waut{t} - \Waut{\infty}}\\
		+ \norm[2]{\Win{t} - \Win{\infty}}.
	\end{multlined}
\end{gather}
We first bound $\norm[2]{\Waut{t} - \Waut{\infty}}$.
For $t\ge1$,
we factor out the common scalar factor $\prodlamfin{0}{t-1}$:
\begin{equation}\label{eq:leg-weights-prod-fact-t>tf}
	\Waut{t} - \Waut{\infty} = \lr \Wleg^{t} - \one v^\top\prodlamfin{t}{\infty}\rr \prodlamfin{0}{t-1}.
\end{equation}
Using the triangle inequality,
we bound the first factor as
\begin{equation}\label{eq:w-true-diff-triangle}
	\begin{aligned}
		\norm[2]{\Wleg^{t} - \one v^\top\prodlamfin{t}{\infty}} 
		&= \norm[2]{\Wleg^{t} - \one v^\top + \one v^\top - \one v^\top\prodlamfin{t}{\infty}}\\
		&\le 
		\norm[2]{\Wleg^{t} - \one v^\top} 
		+ \norm[2]{\one v^\top - \one v^\top\prodlamfin{t}{\infty}}\\
		&= \sigmam^t + 1 - \prodlamfin{t}{\infty},
	\end{aligned}
\end{equation}
where $\norm[2]{\one v^\top}=1$ and we use the singular value decomposition of the doubly stochastic matrix $W$.
Combining~\eqref{eq:leg-weights-prod-fact-t>tf} with~\eqref{eq:w-true-diff-triangle} yields
\begin{equation}\label{eq:leg-weights-prod-norm-bound-t>tf}
	\norm[2]{\Waut{t}-\Waut{\infty}} \le \prodlamfin{0}{t-1}\lr \sigmam^{t} +  1-\prodlamfin{t}{\infty} \rr.
\end{equation}
We now bound $\norm[2]{\Win{t} - \Win{\infty}}$.
Defining, for $s\le t$,
\begin{equation}
	C_{s}^{t} \doteq \prodlamfin{s+1}{t}\Wleg^{t-s} - \one v^\top\prodlamfin{s+1}{\infty},
\end{equation}
we use~\eqref{eq:limits-W} and rewrite $\Win{t} - \Win{\infty}$ as
\begin{equation}\label{eq:A-B-inf-decomp}
	\Win{t} - \Win{\infty}	= \sum_{k=0}^{t-1}C_k^{t-1}\lam{k} - \one v^\top\sum_{k=t}^{\infty}\prodlamfin{k+1}{\infty}\lam{k}.
\end{equation}
The induced norm of the first summation in~\eqref{eq:A-B-inf-decomp} can be upper bounded using the triangle inequality:
\begin{equation}\label{eq:C-norm-sum}
	\norm[2]{\sum_{k=0}^{t-1}C_k^{t-1}\lam{k}} \le \sum_{k=0}^{t-1}\norm[2]{C_k^{t-1}}\lam{k}.
\end{equation}
For each element in the summation in~\eqref{eq:C-norm-sum},
the same argument used to bound $\norm[2]{\Waut{t} - \Waut{\infty}}$ applies,
with the difference that the products start from $k+1$:
\begin{equation}
	\norm[2]{C_k^{t-1}} = \norm[2]{\Wleg^{t-1-k} - \one v^\top\prodlamfin{t}{\infty}}\prodlamfin{k+1}{t-1}
\end{equation}
and,
analogously to~\eqref{eq:w-true-diff-triangle},
we derive
\begin{equation}
	\norm[2]{\Wleg^{t-1-k} - \one v^\top\prodlamfin{t}{\infty}} \le
	\sigmam^{t-1-k} +  1-\prodlamfin{t}{\infty}
\end{equation}
\begin{equation}\label{eq:bound-C}
	\norm[2]{C_k^{t-1}} \le \prodlamfin{k+1}{t-1}\lr \sigmam^{t-1-k} +  1-\prodlamfin{t}{\infty} \rr.
\end{equation}
For the second summation in~\eqref{eq:A-B-inf-decomp},
we readily obtain
\begin{equation}\label{eq:A-second-summation}
	\norm[2]{\one v^\top\sum_{k=t}^{\infty}\prodlamfin{k+1}{\infty}\lam{k}}
	= \sum_{k=t}^{\infty}\prodlamfin{k+1}{\infty}\lam{k}.
\end{equation}
Combining~\eqref{eq:A-B-inf-decomp}--\eqref{eq:A-second-summation} yields
\begin{multline}\label{eq:leg-weights-sum-norm-bound-t>tf}
	\norm[2]{\Win{t} - \Win{\infty}} \le \sum_{k=0}^{t-1}\prodlamfin{k+1}{t-1}\lam{k}\lr\sigmam^{t-1-k} +  1-\prodlamfin{t}{\infty}\rr \\
	+ \sum_{k=t}^{\infty}\prodlamfin{k+1}{\infty}\lam{k}.
\end{multline}
\revision{The assumption $\state{0}\not\parallel\one$ implies that $\norm{\state{0}-\statess\one}\neq0$.}
Hence,
subbing~\eqref{eq:state-mismatch-leg-norm},~\eqref{eq:leg-weights-prod-norm-bound-t>tf},
and~\eqref{eq:leg-weights-sum-norm-bound-t>tf} into~\eqref{eq:state-norm-ineq} yields $\bar{\rho}(t)$ in~\eqref{eq:convergence-rate-upper-bound}.

\subsection{Proof of \Cref{prop:convergence-rate-lower-bound}}
\label{ass:proof-convergence-rate-lower-bound}

Because $(\Waut{t} + \Win{t})\one = \one \ \forall t\ge0$,
it holds
\begin{equation}
	\state{t} - \statess\one 
	= (\Waut{t} + \Win{t})(\state{0} - \statess\one).
\end{equation}
Let $u_2$ and $v_2$ be the orthonormal singular vectors such that $\Wleg v_2 = \sigmam u_2$.
Let $\state{0}$ be such that $\state{0} = \statess\one  + v_2$.
It holds
\begin{equation}
	\begin{aligned}
		\state{t} - \statess\one 
		&= (\Waut{t} + \Win{t})v_2 \\
		&= \prodlamfin{0}{t-1}\Wleg^{t}v_2 + \sum_{k=0}^{t-1} \prodlamfin{k+1}{t-1}\lam{k}\Wleg^{t-1-k}v_2 \\
		&= \prodlamfin{0}{t-1}\sigmam^t u_2 + \sum_{k=0}^{t-1} \prodlamfin{k+1}{t-1}\sigmam^{t-1-k}u_2
	\end{aligned}
\end{equation}
and we obtain the lower bound in~\eqref{eq:conv-rate-lower-bound} as
\begin{equation}
	\begin{aligned}
		\dfrac{\norm[2]{\state{t} - \statess\one}}{\norm[2]{\state{0} - \statess\one}} 
		&= \left| \prodlamfin{0}{t-1}\sigmam^t + \sum_{k=0}^{t-1} \prodlamfin{k+1}{t-1}\sigmam^{t-1-k} \right| \dfrac{\norm[2]{u_2}}{\norm[2]{v_2}}\\
		&= \prodlamfin{0}{t-1}\sigmam^t + \sum_{k=0}^{t-1} \prodlamfin{k+1}{t-1}\sigmam^{t-1-k}.
	\end{aligned}
\end{equation}

\subsection{Proof of \Cref{cor:nonuniform}}
\label{app:nonuniform}

We construct a simple sequence $\{\lamv{t}\}_{t\ge0}$ that satisfies the claim.
Without loss of generality, assume that $\state[1]{0}$, 
(\ie the initial state of the agent labeled as $1$), 
is the (not necessarily unique) largest element of $\state{0}$. 
Let $\state{t}$ denote the state when the agents are following the regular consensus dynamics, 
namely Eq.~\eqref{eq:fj} with $\lamv{t} \equiv 0$,
then since $\state{0} \not\parallel \one$ and $W$ is primitive there exists a time step $t^*$ such that $\state[1]{t} < \state[1]{0} \ \forall \ t \geq t^*$.
\revision{Let $y_{t}$ denote the state trajectory obtained by
using the protocol~\eqref{eq:fj_nonuniform} starting from $\revision{y_0} = \state{0}$ and with $\lamv{t}$ defined as follows:}
\begin{equation}\label{eq:nonuniform-lambda-def}
    \lam[i]{t} = \begin{cases}
        1 & \text{if } i = 1, \ t \leq t^*, \\
        0 & \text{otherwise}.
    \end{cases}
\end{equation}
Then,
it holds $\state[i]{t} \leq y_{t}^i \ \forall \, t,i$ and specifically $\state[1]{t^*} < y_{t^*}^1 = \state[1]{0}$.
For $t>t^*$,
$\lamv{t} = 0$ and the dynamic~\eqref{eq:fj_nonuniform} with~\eqref{eq:nonuniform-lambda-def} simplifies back to the 
consensus protocol.
Then, we have that:
\begin{equation}
    \begin{aligned}
        \lim_{t\rightarrow\infty}\state{t}&=\one v^\top\state{t^*} = \one v^\top \state{0} = \statess\one\\
        \lim_{t\rightarrow\infty}y_t&=\one v^\top y_{t^*}.
    \end{aligned}
\end{equation}
Since the Perron eigenvector $v > 0$, it holds $\lim_{t\rightarrow\infty}\state[i]{t} < \lim_{t\rightarrow\infty}y_{t}^i$ for every $i$ and thus $\lim_{t\rightarrow\infty}y_{t} \neq \statess\one$.
    
    

\end{document}